# Mobile Broadband Possibilities considering the Arrival of IEEE 802.16m & LTE with an Emphasis on South Asia

Nafiz Imtiaz Bin Hamid[1], Md. Zakir Hossain[2], Md. R. H. Khandokar[3], Taskin Jamal[4], Md.A. Shoeb[5]

Department of Electrical and Electronic Engineering (EEE)
[1] Islamic University of Technology (IUT), Board Bazar, Gazipur-1704, Bangladesh.
Email: nafiz.imtiaz@ieee.org
[4] The University of Asia Pacific (UAP), Dhanmondi R/A, Dhaka-1209, Bangladesh.
[5] Stamford University, Siddeswari, Dhaka-1217, Bangladesh.
[3] School of Engineering and Computer Science. Independent University, Bangladesh.
[2] Radio Access Network (RAN) Department, Qubee. Augure Wireless Broadband Bangladesh Limited.

*Abstract*— **This paper intends to look deeper into finding an ideal mobile broadband solution. Special stress has been put in the South Asian region through some comparative analysis. Proving their competency in numerous aspects, WiMAX and LTE already have already made a strong position in telecommunication industry. Both WiMAX and LTE are 4G technologies designed to move data rather than voice having IP networks based on OFDM technology. So, they aren't like typical technological rivals as of GSM and CDMA. But still a gesture of hostility seems to outburst long before the stable commercial launch of LTE. In this paper various aspects of WiMAX and LTE for deployment have been analyzed. Again, we tried to make every possible consideration with respect to south Asia i.e. how mass people of this region may be benefited. As a result, it might be regarded as a good source in case of making major BWA deployment decisions in this region. Besides these, it also opens the path for further research and in depth thinking in this issue.**

## I. INTRODUCTION

Broadband wireless is a technological confluence in bringing the broadband experience to a wireless context, which offers users certain unique benefits and convenience. Broadband is going mass market while consumers are increasingly mobile. Mobile telephony leads the telecommunications market and data services represent an increasing share of total mobile ARPU.Thus, there is a crying need for deploying a cost-effective and scalable wireless broadband technology in this region to meet the broadband hunger of the classes as well as the masses [1].

WiMAX stands for Worldwide Interoperability for Microwave Access. It is a 4th generation cellular telecommunication technology currently based on IEEE 802.16e standard. Mobile WiMAX based on IEEE 802.16e-2005 [2] standard is an amendment of IEEE STD 802.16-2004 [3] for supporting mobility [28] [29]. IEEE 802.16-2004 is also frequently referred to as "Fixed WiMAX" since it has no support for mobility [3].The IEEE 802.16m [4] standard is the core technology for the proposed Mobile WiMAX Release 2, which enables more efficient, faster, and more converged data communications.

Long Term Evolution (LTE) offers a superior user experience along with a simpler technology for next-generation mobile broadband. LTE is the next major step in mobile radio communications and is introduced in 3GPP (3rd Generation Partnership Project) Release 8. It is the last step toward the 4th generation (4G) of radio technologies designed to increase the capacity and speed of mobile telephone networks [5].The world's first publicly available LTE-service was opened by TeliaSonera in the two Scandinavian capitals Stockholm and Oslo on the 14th of December 2009 [6][7][30].

South Asia, also known as Southern Asia, is the southern region of the Asian continent typically consisting of Bangladesh, Bhutan, India, the Maldives, Nepal, Pakistan and Sri Lanka. Some definitions may also include Afghanistan, Myanmar, Tibet, and the British Indian Ocean Territories. South Asia is home to well over one fifth of the world's population, making it both the most populous and most densely populated geographical region in the world [8].So, finding an ideal mobile broadband solution for the mass population in this region is really a vital decision and thus requires analyzing from various point of view.

In this paper, an organized attempt has been made to facilitate a planned decision making stage for the mobile broadband solution in this region.

## II. MOBILE BROADBAND – AN OVERVIEW

There are strong evidences supporting predictions of increased mobile broadband usage. Consumers understand and appreciate the benefits of mobile broadband. Broadband subscriptions are expected to reach 3.4 billion by 2014 and about 80 percent of these consumers will use mobile broadband shown in Figure 1 [9].

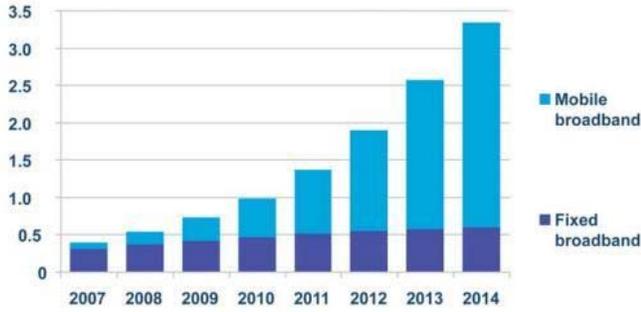

Figure 1.   Fixed and mobile broadband growth

Figure 2 shows a comparative analysis of mobile broadband subscribers per 100 households [31]. It says mobile broadband penetration so far is 13.2% in developed countries, 0.8% in developing countries and on an average it is 3% all over the world.

Mobile broadband is becoming a reality, as the internet generation grows accustomed to having broadband access wherever they go and not just at home or in the office.

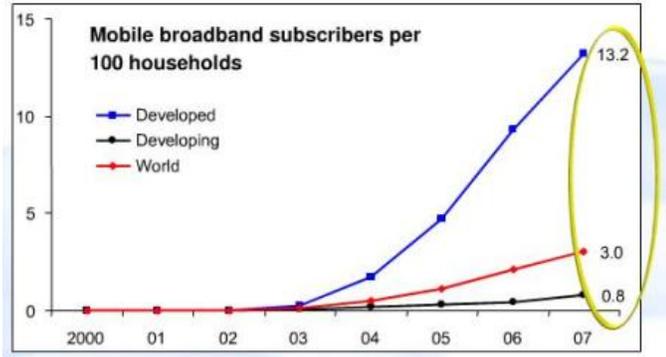

Figure 2.   Mobile Broadband Subscribers:A Comparison

## III.   WHY LTE & WiMAX SHOULD COME INTO FOCUS

There are many technology candidates for mobile broadband. As a result, there remains an intense competition. But there are some distinct factors for determining technology choices; namely-ecosystem, regulation, supporting lobbies, standardization and technical aspects. Five technologies namely-TD-CDMA, HSPA, EVDO, WiMAX & LTE can be chosen as the satisfactory ones. But WiMAX and LTE far exceed all others considering these determining issues [32]. Logically so, in the pathway for choosing an ideal mobile broadband solution; we will focus on WiMAX with its existing standard IEEE 802.16e & LTE. Again, decision making scenario will possibly undergo a severe complexity when WiMAX 2 will come into play with its upgrade IEEE

802.16m.It is expected to offer similar performance like LTE.LTE will also have its upgrade named as LTE-Advanced. Our analysis will try to integrate all these interesting but critical aspects to find a reliable solution.

## IV.   WiMAX AND LTE:TECHNICAL ASPECTS

Current WiMAX network uses IEEE standard 802.16e for ensuring mobility. Future ratification of the current standard is IEEE 802.16m with improved features and higher data rates where the network update procedure is also possible from the current one [4] [8] [18] [19]. LTE will take time to roll out, with deployments reaching mass adoption by 2012 while WiMAX is out right now. The table presented below [17] shows a comparative overview between these popular standards IEEE 802.16e & 3GPP-LTE.

TABLE I.        TECHNICAL OVERVIEW OF IEEE 802.16E & LTE

| Aspect | Mobile WiMAX (IEEE802.16e-2005) | 3GPP-LTE (E-UTRAN) |
|---|---|---|
| Core network | WiMAX Forum™ All-IP network | UTRAN moving towards All-IP Evolved UTRA CN with IMS |
| Access technology: Downlink (DL) Uplink (UL) | OFDMA OFDMA | OFDMA SC-FDMA |
| Frequency band | 2.3-2.4GHz, 2.496-2.69GHz, 3.3-3.8GHz | Existing and new frequency bands (~2GHz) |
| Bit-rate/Site: DL UL | 75Mbps (MIMO 2TX 2RX) 25Mbps | 100Mbps (MIMO 2TX 2RX) 50Mbps |
| Channel bandwidth | 5, 8.75, 10MHz | 1.25-20MHz |
| Cell radius | 2-7km | 5Km |
| Cell capacity | 100-200 users | >200 users @ 5MHz >400 users for larger BW |
| Spectral efficiency | 3.75[bits/sec/Hz] | 5[bits/sec/Hz] |
| Mobility: Speed Handovers | Up to 120Km/H Optimized hard handovers supported | Up to 250Km/H Inter-cell soft handovers supported |
| Legacy | IEEE802.16a through 16d | GSM/GPRS/EGPRS/UMTS/HSPA |
| MIMO: DL UL No. of code words | 2Tx X 2Rx 1Tx X NRx (Collaborative SM) 1 | 2Tx X 2Rx 2Tx X 2Rx 2 |
| Standardization coverage | IEEE 802.16e-2005 PHY and MAC CN standardization in WiMAX forum™ | RAN (PHY+MAC) + CN |
| Roaming framework | New (work in process in WiMAX Forum™) | Auto through existing GSM/UMTS |
| Schedule forecast: Standard completed Initial Deployment Mass market | 2005 2007 through 2008 2009 | 2007 2010 2012 |

## V.   FUTURE UPGRADES TO APPEAR IN STANDARDS

Future upgrade of current WiMAX standard IEEE 802.16e is IEEE 802.16m whether for LTE it will be LTE-Advanced. Both these upgrades will substantially enhance the performance of the previous standards. Thus, existing mobile broadband scenario will experience potential change and consumers will relish better mobile broadband.

*A. IEEE 802.16m*

The IEEE 802.16m standard is the core technology for the proposed Mobile WiMAX Release 2 [8]. Among many enhancements, IEEE 802.16m systems can provide four times faster data speed than the current Mobile WiMAX Release 1 based on IEEE 802.16e technology. It can push data transfer speeds up to 1 Gbit/s while maintaining backwards compatibility with existing WiMAX radios [4][8]. IEEE 802.16m shall meet the cellular layer requirements of IMT-Advanced next generation mobile networks [20].Table II shown below [20] presents the proposed Air Interface Requirement for this IEEE 802.16m standard.

The 802.16m profile is currently under evaluation and is expected to be ratified along with WiMAX Release 2 later this year. The first 802.16m dongles is expected to be visible in late 2011 while more wide-spread commercial deployments is to be starting in 2012 [21].

TABLE II. AIR INTERFACE REQUIREMENT FOR IEEE 802.16M

| Characteristic | | Requirement |
|---|---|---|
| Peak Data Rate | Downlink | 1Gbps @ Nomadic<br>100Mbps @ High mobility |
| | Uplink | TBD |
| Expected spectral efficiency | Micro cell (DL/UL) | TBD |
| | Macro cell (DL/UL) | TBD |
| Bandwidth | | Scalable bandwidth including 5, 7, 8.75, 10 MHz |
| Center frequency | | Frequency is expected to be decided in WRC07 |

*B. LTE-Advanced*

LTE-Advanced extends the technological principles behind LTE into a further step change in data rates. Incorporating higher order MIMO (4x4 and beyond) and allowing multiple carriers to be bonded together into a single stream, target peak data rates of 1Gbps have been set. It also intends to use a number of further innovations including the ability to use non-contiguous frequency ranges, with the intent that this will alleviate frequency range issues in an increasingly crowded spectrum, self back-hauling base station and full incorporation of Femto cells using Self-Organizing Network techniques [25].

TABLE III. LTE-ADVANCED WITH SOME MAJOR AIMS

| Peak data rates | DL:1Gbps<br>UL:500Mbps |
|---|---|
| Spectrum efficiency | 3 times greater than LTE |
| Peak spectrum efficiency | DL:30 bps/Hz<br>UL:15 bps/Hz |
| Spectrum use | The ability to support scalable bandwidth use and spectrum aggregation where non-contiguous spectrum needs to be used |
| Latency | From Idle to Connected in less than 50 ms and then shorter than 5 ms one way for individual packet transmission |
| Cell edge user throughput | Twice that of LTE |
| Average user throughput | 3 times that of LTE |
| Mobility | Same as that in LTE |
| Compatibility | LTE Advanced shall be capable of interworking with LTE and 3GPP legacy systems |

Some of the main aims for LTE-Advanced are shown [26] in Table III.

❖ **LTE-Advanced in comparison with other 3G competitors:**

The development of LTE Advanced/IMT Advanced can be seen as an evolution from the 3G services that were developed using UMTS / W-CDMA technology. Compared to other 3G services like UMTS, HSPA, HSPA+; LTE is far ahead. So, it goes without saying that LTE-Advanced should obviously be a dazzler in carrying on the superiority. The following table shows the comparative performance analysis with respect to some vital parameters:

TABLE IV. LTE-ADVANCED COMPARED TO LTE AND OTHER 3G TECHNOLOGIES

| | WCDMA (UMTS) | HSPA HSDPA / HSUPA | HSPA+ | LTE | LTE ADVANCED (IMT ADVANCED) |
|---|---|---|---|---|---|
| Max downlink speed bps | 384 k | 14 M | 28 M | 100M | 1G |
| Max uplink speed bps | 128 k | 5.7 M | 11 M | 50 M | 500 M |
| Latency round trip time approx | 150 ms | 100 ms | 50ms (max) | ~10 ms | less than 5 ms |
| 3GPP releases | Rel 99/4 | Rel 5 / 6 | Rel 7 | Rel 8 | Rel 10 |
| Approx years of initial roll out | 2003 / 4 | 2005 / 6 HSDPA 2007 / 8 HSUPA | 2008 / 9 | 2009 / 10 | |
| Access methodology | CDMA | CDMA | CDMA | OFDMA / SC-FDMA | OFDMA / SC-FDMA |

## VI. RELATED WORKS

One of the key missions of this paper is to emphasize the region South Asia while traveling towards the mentioned destination. That is we approached for finding a suitable broadband solution but regional consideration always obtained a vital significance. South Asia though having an enormous population is still considered as a neglected, developing region. So, works based on South Asia is relatively less. In case of finding suitable mobile broadband solution we couldn't find any related work with the best of our effort. But it is of great importance as it may change the fate of mass population in this region.

The IEEE standards for fixed and mobile WiMAX remain consecutively in [2] & [3]. The technical overview of IEEE standard 802.16m, which is supposed to work in parallel with mobile WiMAX release 2 are given in [4],[18],[20],[21] and [22].Various significant technical aspects of WiMAX and LTE have been given in the books referenced as [1],[5],[7] & [37].A suitable mobility handover scenario has been depicted in [28] & [29]. Analysis of the general notion about the WiMAX and LTE and obstacles behind their proper flourishment have been elaborated in [10],[11],[17],[19] and [32] .They also try to break the wrong and inimical interrelations between these two technologies.

Other references like those from [30],[31],[33]-[36] reflect various up to date technological news in case of network deployment or device arrivals. Those have been picked in various sections of this paper to strengthen the overall analysis.

To the best of our knowledge, we couldn't get with in touch with such sort of analysis methodology used by us in this paper.

## VII. PRESENT NETWORK DEPLOYMENT SCENARIO

Although the hype around WiMAX is quickly dissipating, still the standard has gained enough backing and volume to serve as an alternative for the provisioning of mobile broadband access. It has begun to carve out a tight niche tied to certain target opportunities, it has inspired a new wireless business model, and it has a flexible, flat, all-IP network architecture better suited than HSPA to providing Internet-based services. In contrast, however, the LTE standard has quickly gained substantial momentum. Since WiMAX 802.16e and LTE release 8 will provide similar real-world performance, ultimately the decisions of the largest WiMAX players may determine the fate of WiMAX [10].

The number of WiMAX deployments — currently more than 500 across 145 countries is  greater than that of any conventional 3G technology and more than 50% greater than the  number of HSPA network commitments. However, most WiMAX deployments to date have been small and expected to increase the coverage. Many of the larger WiMAX deployments are still underway, and large countries such as India, Indonesia and Vietnam are just beginning to issue WiMAX licenses [10].

TelecomAsia recently reported in Asia Pacific there were 55 LTE networks in these categories compared to 25 mobile WiMAX networks in 1st quarter of 2009, although mobile WiMAX had the clear advantage of 12 networks in service compared to zero for LTE, according to Informa Telecoms & Media research.[11].The fact is shown below [33] in Figure 3.

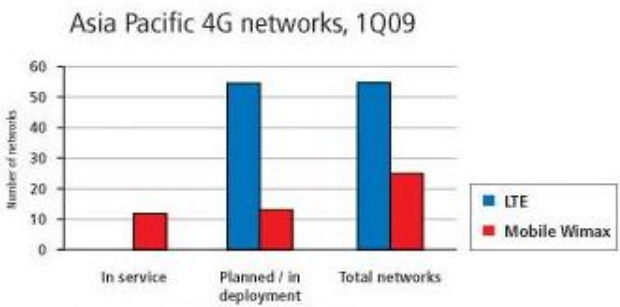

Figure 3.   Comparison of LTE and WiMAX in Asia Pacific 4G Networks

## VIII. DEPLOYMENT IN SOUTH ASIA-A CASE STUDY

South Asia is home to well over one fifth of the world's population. Being such a densely populated developing region in the world map; the answer of ideal mobile broadband solution for this zone is quite a challenging case. There remains a huge difference in the life style led by various classes of people of this region. So, when the question of choosing the perfect mobile broadband solution comes. It must have to be economically suitable to the general people. Let's now take a quick look on the current status of different south Asian countries in case of mobile broadband deployment:

### A. WiMAX Network

#### 1) Bangladesh:

Agni- A Public Limited Company started providing Internet Access in Bangladesh since 1995; launched Pre-WiMAX in 2006. Later on they started deploying Fixed WiMAX on 3.5 GHz Band before the arrival of mobile broadband. Then the era of mobile broadband started with Qubee from Augere with a mission "Broadband for all". They launched wireless broadband internet services for residential and business customers in Dhaka in October 2009 [8] [12]. Banglalion acquiring the 1st BWA license from Bangladesh Telecommunication Regulatory Commission (BTRC) on November 18, 2008 and committed to provide Broadband Internet connectivity and other services using WiMAX technology. By the 2nd quarter of 2010, Banglalion plans to bring the whole Dhaka city and it's wider periphery under WiMAX coverage. By the end of 2nd quarter 2011, Banglalion plans to bring most areas of the country under its seamless coverage [13].

#### 2) India:

Though India does not have that many notable B2C WiMAX networks, a few companies are doing on experimental basis.BSNL,India haslaunched a commercial network in Kerala in late February,2010, which is the first urban commercial 4G network in India[34].They did also have success in network deployment during July,2009 in Goa along with Gujrat. They are also aiming for Punjab quite soon.BSNL has 20 MHz of BWA spectrum in 2.5 Ghz frequency.They already completed the initial deployment of 1000 BS and planning to launch more quickly [8][14].Except BSNL, other contributions came through Aircell through its Business Solution (ABS) plan .It is the first company in India to launch WiMAX and one among the five global operators to achieve this feat [8].They initiated their dominance through WiMax-ing Chennai. Again, Reliance launched servics on Pune,Bangalore on July 4,2009 and VSNL Tata Indicom is also in the way of launching service [8].

#### 3) Pakistan:

Wi-Tribe Pakistan bringing together two shareholders Qatar Telecom (Qtel) and Saudi-based A.A started commercial services in June 2009 in 4 major cities, while expansion is underway for another 10 cities.Qubee has launched services for residential and business customers in Karachi. Super Broadband service is launching soon with testing network running in Karachi. Mobilink Infinity Pakistan's Second WiMAX network has operations in the South region currently [8]. Wateen Telecom headquartered in Lahore has successfully deployed one of the biggest nationwide WiMAX networks with 42 MHz of spectrum. It started commercial launch in December,2007; now deployed more than 842 four sector base stations across 22 cities with network covering 20% of Pakistan's 164 million inhabitants [8] [14].

### 4) Srilanka:

In Srilanka, Dialog Telekom began commercial operations in late 2006 offering speeds of up to 4 Mbit/s. Sri Lanka Telecom has also launched test transmission in certain areas.Lanka Bell started commercial operations in early 2008.Again,SUNTEL also has started the WiMAX Broadband [8].

### B. LTE Network Deployment

In December 2009, TeliaSonera became the first operator in the world to offer commercial 4G/LTE services, in the central parts of Stockholm and Oslo; network being supplied by Ericsson and Huawei [15] [8].

But reality for South Asia isn't the same. Potential network opening time for this region is way too far from the above mentioned one. Up to now in the South Asian region only India and Pakistan have some definite plan for LTE deployment. But it isn't before quarter 4 of 2012.

The following table shows the deployment plan for the south Asian countries for possible deployment of LTE [16].

TABLE V.     POSSIBLE DEPLOYMENT PLAN OF LTE IN SOUTH ASIA

| Country | Operator | Potential Opening Time |
|---------|----------|------------------------|
| India | BSNL | Q4 2012 |
| Pakistan | PMCL | December 2014 |
| | Telenor | March 2014 |

As a result, scope of reconsidering any sort of mass deployment or any strategic change in this region is still possible.

### IX.   ARRIVAL OF IEEE 802.16M,LTE-ADVANCED & THEIR IMPACTS

### A. Backward Compatibility

According to the announcement of WiMAX Forum, WiMAX Release 2, which is based on the standard, would be finalized in parallel with 802.16m.It is just to ensure that the next generation of WiMAX networks and devices will remain backward compatible with WiMAX networks based on 802.16e [21].

Samsung's Mobile WiMAX solution provides strong backward compatibility with Mobile WiMAX Release 1 solutions. It allows current Mobile WiMAX operators to migrate their Release 1 solutions to Release 2 with little expense by upgrading channel cards and their base station software. Also, the subscribers who use currently available Mobile WiMAX devices can communicate with new Mobile WiMAX Release 2 systems without difficulty [22].

LTE has the advantage of being backwards compatible with existing GSM and HSPA networks, enabling mobile operators deploying LTE to continue to provide a seamless service across LTE and existing deployed networks [25].

### B. Spectrum Issue

Most 3G networks operate using up to 5MHz channels, WiMAX 802.16e networks operate using up to 10MHz, and 802.16m and LTE networks will operate using up to 20MHz channels. To achieve the significantly higher performance as reported by TeliaSonera, LTE operators need to use the wider 20MHz channels, but that spectrum is not always readily available [21].Lots of the spectrum allocation are in 10MHz chunks and the places with contiguous 20MHz channels are few and far between according to Intel.

So, this spectrum issue might add to complexity for the overall satisfactory performance of LTE.LTE-Advanced bearing 3GPP Release 10 may bring some relief in this regard.

### X.   DEPLOYMENT SIGNIFICANCE ON REGIONAL BASIS

There must have be some regional benefits in South Asia for successful mobile broadband deployment. It must have to bring significant changes in the lifestyle of the mass people of this region. Otherwise that deployment will obviously fall under severe questionnaire. Some significant factors are given below which tell about the possible betterment through successful mobile broadband deployment.

### A. Disaster management

WiMAX access was used to assist with communications in Aceh, Indonesia, after the tsunami in December 2004 when all communication infrastructure in the area, other than amateur radio, was destroyed, making the survivors unable to communicate with people outside the disaster area and vice versa. Then WiMAX provided broadband access that helped regenerate communication to and from Aceh.

In addition, WiMAX was donated by Intel Corporation to assist the FCC and FEMA in their communications efforts in the areas affected by Hurricane Katrina. In practice, volunteers used mainly self-healing mesh, VoIP, and a satellite uplink combined with Wi-Fi on the local link [4].

In the South Asian region, we often find the natural calamities which badly affect people. Then the sufferings of poor people know no bound. For the disaster management plan mobile broadband solution like WiMAX can contribute a greater role if it can be deployed successfully all over the country.

❖   Seismic Activity Detection:
Seismic activity has an impact on the ground shaking and rupture and also responsible for other exclusive natural disasters such as tsunamis, floods, and fires. Seismic wave can be detected using several sensors connected through multifarious facilitated WiMAX or LTE Network.

Meteorological department of a country normally works under the administrative control of Ministry of Defense. It is mainly responsible for recording the metrological and seismological observations and warnings for disaster management. The mission is to ensure improved protection of life, property and environment; increased safety on land, at sea and in the air; enhanced quality of life and sustainable economic growth [36]. Figure 4 shown below represents the seismic communication network of Bangladesh Meteorological Department (BMD).

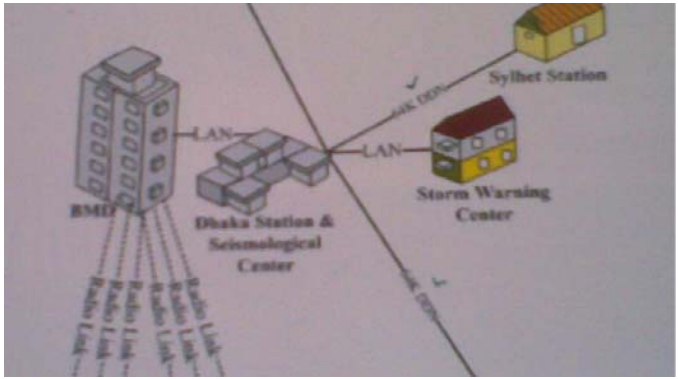

Figure 4. Seismic Communication Network of BMD

♦ WiMax may substantially improve existing monitoring systems by enabling wireless broadband connectivity to remote sites; mobile access, and the deployment of a more comprehensive WAN system.

♦ Again, it is worth mentioning that network can be quite enormously diversified through the use of mobile broadband technology like mobile WiMAX with reduced wire connection.

♦ Improvement of the focused integrated seismic network in regions by connecting the unconnected sensors via chipset/modem or handheld device.

♦ LTE-Advanced even LTE can give birth to a miracle in this respect

*B. Telemedicine*

Video Conferencing has become a vital component in all human activities. Telemedicine System is not apart in this field. There are video conferencing systems that allow doctors from different hospitals to communicate with minimum requirements such as a Webcam and of course an Internet connection. One such example is the eBaithak-a real time multiparty video conferencing System developed by IIT Kharagpur [35].

But ensuring mobility in video conferencing is always a challenge from the network point of view. That is in an emergency condition one of the parties can be in a vehicle. Figure 5 gives the idea of a typical idea of remote assistance via telemedicine. But the efficacy of this video conference depends on the network that can successfully tackle the handover.

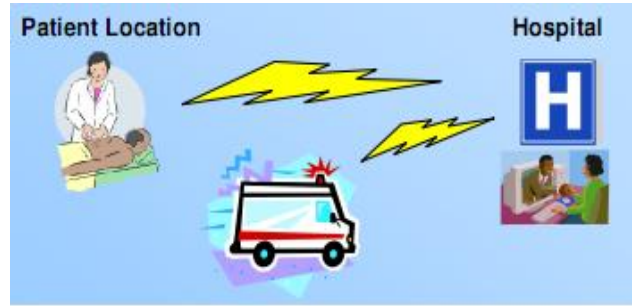

Figure 5. Remote assistance through Telemedicine

♦ Mobile WiMAX with its scheduling service i.e. Real Time Polling Service (rtPS), Enhanced Real Time Polling Service (ertPS) which support audio and video streaming and VOIP with activity detection can resolve this scenario [2].

♦ In fact, ertPS supports IEEE 802.16e for VOIP support with variable packet size with silence suppression [2].

♦ Things will get even brighter with the arrival of IEEE 802.16m with WiMAX release 2 & LTE associated by its upgrades.

*C. Traffic Control with Vehicle Tracking*

The availability of an all IP, mobile, high-speed network has also generated some innovative applications. To better help students track the location of shuttle buses, Ball State university- a cutting-edge wireless research university in Muncie, Indiana; installed notebook computers equipped with WiMAX and GPS USB dongles in the front panels of the busses. The information is then reported back to a fleet management server, providing real-time location based information that can be accessed by the students [23].

▪ South Asia is enormously being affected by poor traffic control system. In this region, to ensure a happy organized life of people maintaining a planned traffic control system is deadly necessary.

▪ But we are finding that lots of plans being initiated are going in vain as some of the cases of Bangladesh. In this regard, we can think of building a real-time location based system employing WiMAX which is the available mobile broadband solution option right now. Later on we can think of utilizing LTE or LTE-Advanced in this regard.

▪ Planned time information based bus/train service with appropriate arrival time for the passengers may be provided with this in this region.

▪ Again, for security issue like car, taxi theft vehicle tracking system is deadly necessary. Mobile broadband with greater QoS like LTE & WiMAX can facilitate this quite efficiently.

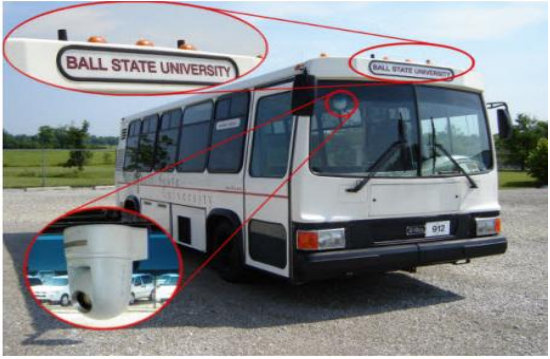

Figure 6.   Providing real-time location based information

Figure 6 depicts such utilization of real-time information based system with WiMAX [23].

### D. Connecting Banking Networks

The banking system where security is the major concern can be connected through the WiMAX networks. Owing to the broad coverage and large connectivity, WiMAX can connect a large number of diversely located banks and ATM locations [37].

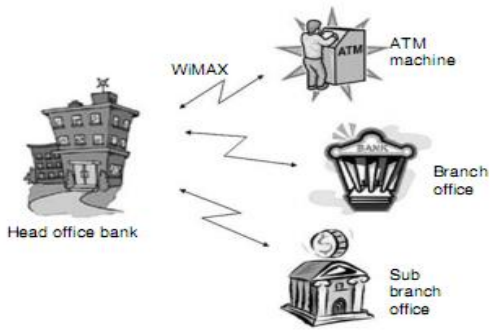

Figure 7.   Banking System Connected by WiMAX

Such a secured scenario may ensure the following features:

- Monitoring and collaboratively assisting  the employee activities

- Checking the disturbance or inconvenience felt by client in ATM booth and providing assistance.

- Giving card holder the partial authorization for network connectivity to check his balance or the state of the near by ATM booths. It can help remove the harassment of rushing into a not working or under maintenance ATM booth.

## XI.   MARKET PENETRATION AND PROBABLE GROWTH

### A. WiMAX

Notebooks and cellular phones are about to be getting WiMAX enabled. The technological era has already begun. New Intel Core Series notebook with integrated Wi-Fi/WiMAX module can easily facilitate most of the previously mentioned application challenges issues [24]. Again, cellular phone with GSM/WiMAX dual mode is about to come; it also can smooth the scenario towards more improvement. HTC already announced the HTC MAX 4G device that is the world's first GSM/WiMAX headset [27].

### B. LTE

Samsung is working as a technological pioneer in WiMAX release 2 for 4G development. But they aren't just prepping WiMAX release 2 (802.16m) for 4G deployment but also working on developing LTE USB modem. Samsung GT-B3710, the First LTE USB Modem Achieved Interoperability from Ericsson LTE Network in Stockholm. LG is also working dominantly in this respect by recently unveiling its LTE USB modem at CommunicAsia 2009 and showing up another model at CEATEC JAPAN 2009 [6].

## XII.   SUGGESTIONS TO MEET FUTURE CHALLENGES

### A. Key Findings

- WiMAX and LTE aren't the technological rivals like GSM and CDMA. Around 80% of the applied technologies in LTE and WIMAX are almost same.

- With the future upgrade of current standard of WiMAX IEEE 802.16e to IEEE 802.16m; LTE and WiMAX will tend to provide quite similar features.

-  It can't be denied that WiMAX has got a certain advantage of earlier deployment in this wireless field. Still LTE will start in South Asia most possibly at the end of 2012 [16] with attractive options. Again, it has also an excellent update i.e. 3GPP Release 10, LTE-Advanced [26].

- If they are deployed according to their chronological arrival in the telecom industry with no rivalry among the network vendors and operators; we can hope for getting a healthy environment and thus a right broadband solution for subscribers in this region.

- Planned deployment may ensure the arrival of two quite similar technologies based on performance around 2012.It will obviously increase the choice for consumers in quality mobile broadband solution. As a result, mobile broadband experience will possibly be at its level best.

### B. Proposal

In the ideal mobile broadband deployment path the obstacles are that there are some strict 3GPP standard follower vendors and operators. Again, rules imposed by corresponding regulatory commission also do have some significant effects on this issue of deployment decision. To make these decision making as conflict less as possible we hereby propose the following path way of deployment. It is shown in Figure 8 given below.

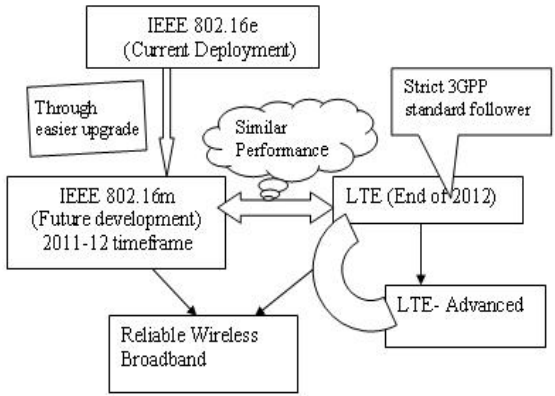

Figure 8. Proposed pathway for conflict less mobile broadband solution

## XIII. CONCLUSION

Irrespective of the regional requirement mobile broadband popularity is increasing day by day. But problem bound South Asian region will surely find it quite difficult to get the best out of it; if inter-vendor or operator conflicts dominate all the deployment decisions. As backward compatibility can be ensured in IEEE 802.16m through easier upgrade from IEEE 802.16e; WiMAX should never lag awkwardly behind LTE or LTE-Advanced. Again, earlier deployment should give it the advantage over LTE. But it shouldn't create the rivalry in between them. Because, the overall internal technologies used by them are mostly same .Now South Asia has WiMAX deployments in some countries which are likely to spread. But LTE deployment should take time and possibly LTE won't enter in South Asia before Q4 2012.But mobile broadband being an enormously popular technology with varied features, use of existing technology i.e. mobile WiMAX right now should be encouraged. Otherwise effective penetration won't take place through the step by step upgrade WiMAX Release 2 (IEEE 802.16m).In the same way then, LTE won't be able to get the dazzling upgrade LTE-Advanced. So, we must have to ensure our level best effort to get the best out of these technologies in various aspects. In this paper these issues have been elaborated suitably with respect to South Asia perspective and the current scenario. Significant utilizations of these technologies like disaster management, secured and suitable banking network, telemedicine, traffic control have been put in light with the regional consideration. Along with all these, a possible prospective solution path towards finding the best possible mobile broadband experience for subscribers has been depicted here.

AUTHORS PROFILE

**Nafiz Imtiaz Bin Hamid** received his Bachelors degree in Electrical and Electronic Engineering from Islamic University of Technology(IUT) in 2008 and now pursuing his M.Sc. Engineering degree.He has been working as a lecturer in the Electrical and Electronic Engineering Department of IUT since 2009.His primary research interest includes Broadband Wireless Access (BWA) i.e. PHY/MAC layer protocol along with mobility related issue analysis of 4G cellular technologies like WiMAX,LTE etc.Nafiz is a graduate student member of IEEE and ACM.He is also a member of IACSIT.Nafiz is included in the Technical Program Committee of ISIEA 2010 & CSSR 2010 to be held in Malaysia.He is also the reciewer of several peer-reviewed International Journals.

**Md. Zakir Hossain** received his Bachelors degree in Electrical and Electronic Engineering from Islamic University of Technology(IUT) in 2008.Currently he is working in the Radio Access Network (RAN) department of QUBEE, Augere Wireless Broadband Bangladesh Limited .

**Md. R. H. Khandokar** received his Bachelors degree in Electrical and Electronic Engineering from Islamic University of Technology(IUT) in 2008.Currently he is serving as a lecturer in the School of Engineering and Computer Science (SECS) of Independent University ,Bangladesh.

**Taskin Jamal** received his Bachelors degree from Electrical and Electronic Engineering department of Islamic University of Technology(IUT) in 2008.Currently he is working as a lecturer in the Electrical and Electronic Engineering department of The University of Asia Pacific (UAP),Bangladesh.

**Md. A. Shoeb** received his Bachelors degree from Electrical and Electronic Engineering department of Islamic University of Technology(IUT) in 2008.He is now pursuing his M.Sc. Engineering degree from Bangladesh University of Engineering and Technology (BUET).Again,he is also serving as the lecturer in Electrical and Electronic Engineering Department of Stamford University,Bangladesh.